\documentclass[
              reprint, twocolumn,
              preprintnumbers,
              amsmath,
              amssymb,
              showpacs,
              superscriptaddress]{revtex4-2}

 \usepackage[utf8x]{inputenc}
 \usepackage{latexsym}
 \usepackage{amsfonts}
 \usepackage{amsmath}
 \usepackage[dvips]{graphicx}
 \usepackage{color}
 \usepackage{bm}
 \usepackage{hyperref}
 \usepackage{lineno}
 \usepackage[autostyle=false]{csquotes}
 \usepackage[most]{tcolorbox}
 \usepackage{verbatim}

\graphicspath{ {./fig/} }
\hypersetup{
   colorlinks=true,
   linkcolor=magenta, 
   citecolor=magenta,
   filecolor=magenta,
   urlcolor=blue   
}
\usepackage{braket}
\usepackage{dcolumn}
\usepackage{enumitem}% http://ctan.org/pkg/enumitem, custom enumerate
\usepackage{mathtools}
\usepackage{color}
\usepackage{xcolor}
\usepackage{float}
\usepackage{bbold}
\usepackage{url}

\newcommand{\ev}[1]{\langle #1 \rangle} % expectation value
\newcommand{\Exe}[0]{(E \times e)}

\begin{document}
\title{Collective Vibronic Cascade in Cavity-Coupled Jahn-Teller Active Molecules
}
\newcommand{\affA}{Department of Chemistry, Indian Institute of Technology Madras, 600036 Chennai, India}
\newcommand{\affB}{Department of Physics, Indian Institute of Technology Madras, 600036 Chennai, India}
\newcommand{\affC}{Center for Quantum Information, Communication and Computing, Indian Institute of Technology Madras, Chennai 600036, India}

\affiliation{\affA}
\affiliation{\affB}
\affiliation{\affC}

\author{Suraj Kumar Pandit}
\thanks{These authors contributed equally to this work.}
\affiliation{\affA}\affiliation{\affC}

\author{Abhinay Pandey}
\thanks{These authors contributed equally to this work.}
\affiliation{\affB}\affiliation{\affC}

\author{Athreya Shankar}
\email{athreya@iitm.ac.in}
\affiliation{\affB}\affiliation{\affC}
\thanks{Corresponding author}

\author{Krishna R. Nandipati}
\email{knandipati@iitm.ac.in}
\affiliation{\affA}\affiliation{\affC}
\thanks{Corresponding author}

\date{\today}

\begin{abstract}
We study the polaritonic states and dynamics of multiple Jahn-Teller (JT) active molecules coupled to the modes of a Fabry-Perot cavity. We find that collective effects dramatically alter the interplay of electronic, vibrational and cavity angular momenta, giving rise to markedly different polaritonic spectra and dynamics even when going from one to two JT molecules. Starting from the ground vibronic state, we find that JT molecules collectively coupled to a common cavity can access a cascade of high-angular-momentum vibronic states in the presence of a single cavity photon, in sharp contrast to the single molecule case where the range of accessible vibronic angular momentum values are bounded. The observable consequences are a broadening of the cavity-molecular polariton spectrum and a suppression of photon polarization dynamics under broadband excitation of the system. Our results uncover new pathways for vibronic angular momentum transfer unique to collective molecular polaritonics with potential implications for cavity-assisted photo-physics and photo-chemistry. 
\end{abstract}

\maketitle

\emph{Introduction.---} Molecular vibronic polaritons are hybrid molecule-cavity quantum systems that offer rich applications in physics and chemistry owing to the mixed electronic, vibrational and photonic character of the system eigenstates~\cite{herrera2020molecular,herrera2016cavity,herrera2017dark,gu2023cavity,gu2021optical,tichauer2022identifying,vendrell2018coherent,vendrell2018collective,polak2020manipulating,kowalewski2016non,ribeiro2018polariton,de2025hybrid,kuppusamy2024spin,GurlekPRR,xiang2024molecular,bhuyan2023rise}.
In particular, vibronic polaritons in highly symmetric organic and inorganic molecular systems, such as Jahn-Teller (JT) active molecules, have been proposed as hybrid platforms for analogue simulators~\cite{de2025hybrid,xiang2024molecular}, quantum computing~\cite{kuppusamy2024spin,xiang2024molecular} and for optomechanical applications~\cite{GurlekPRR,kuppusamy2024spin,de2025hybrid,AntonPhysRevB,xiang2024molecular}. On the chemistry front, vibronic polaritons have been shown to be useful for  controlling excited-state dynamics and spectroscopy~\cite{herrera2020molecular,herrera2016cavity,herrera2017dark,gu2023cavity,gu2021optical,tichauer2022identifying,vendrell2018coherent,vendrell2018collective,polak2020manipulating,kowalewski2016non,xiang2024molecular}, and thereby the photochemistry of molecules~\cite{herrera2020molecular,herrera2016cavity,gu2023cavity,gu2021optical,tichauer2022identifying,vendrell2018coherent,polak2020manipulating,kowalewski2016non,bhuyan2023rise}.  Thanks to the rich internal structure presented by molecules, the interaction of a single molecule with one or more cavity modes is already a problem with a complex interplay of electronic, vibrational and photonic degrees of freedom~\cite{gu2023cavity,gu2021optical,tichauer2022identifying,vendrell2018coherent,kowalewski2016non}. The addition of multiple molecules opens the avenue for an even greater range of \emph{collective} effects, which have been exploited in applications such as controlling chemical reactions via vibrational and electronic strong coupling~\cite{herrera2016cavity,ebbesen2016hybrid,vendrell2018collective,feist2018polaritonic,ahn2023modification,bhuyan2023rise}. Nevertheless, studies of collective effects arising in molecule-cavity systems are frequently limited to empirical observations, with a clear elucidation of vibronic-photonic coupling mechanisms remaining elusive due to the numerous degrees of freedom involved in these systems.

\begin{figure}[t]
  \centering
 \includegraphics[width=0.8\columnwidth]{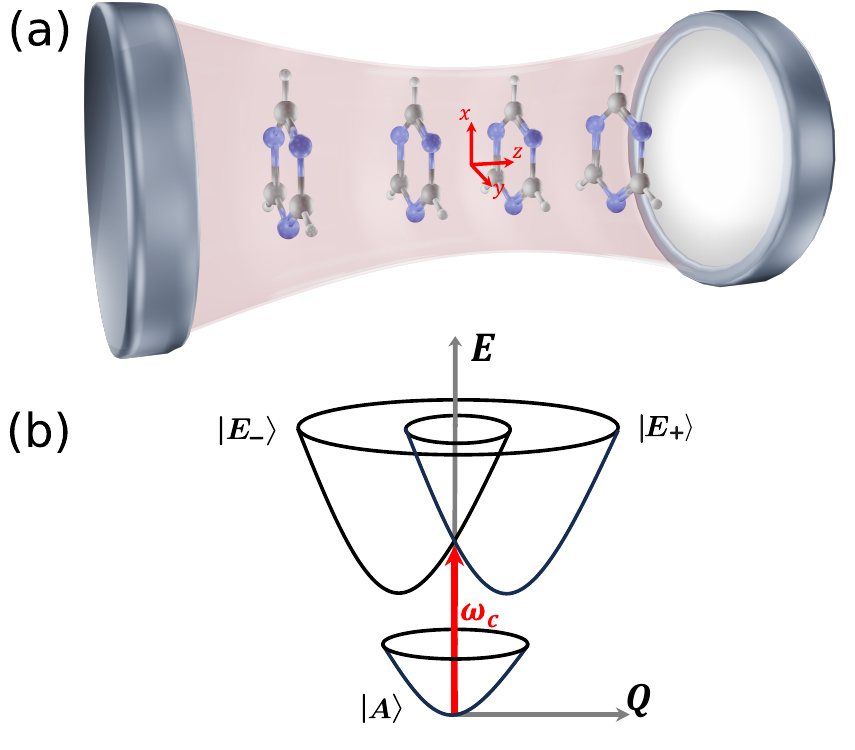}
  \caption{
    (a) Schematic of $\Exe$ Jahn-Teller (JT) active molecules (represented here by sym-triazine systems) situated inside a Fabry-Perot cavity. The symmetry axis of the JT molecules and the wave vector $\vec{k}$ of the $x(y)$-polarized cavity light point along the $z$-axis. (b) JT coupled electronic $\ket{E_\pm}$ states and the ground $\ket{A}$ state of the molecules, sketched as a function of vibrational mode coordinate ($Q$), coupled via the cavity modes with frequency $\omega_c$ (red arrow). Two circularly polarized modes with angular momentum $\pm\hbar$ can be defined from the linearly polarized $x$ and $y$ modes of the cavity~\cite{bar_12_013845}.}
     \label{fig:setup}
\end{figure}

In this Letter, we demonstrate that collective vibronic-photonic coupling leads to an inter-molecular angular momentum transfer mechanism, which we term as \emph{collective vibronic cascade}, in a system consisting of a collection of JT active molecules coupled to a single photon in a Fabry-Perot (FP) cavity. Using a minimal model description in terms of the paradigmatic ($E \times e$) JT Hamiltonian~\cite{longuet1958studies}, we elucidate the mechanism via basis-level analysis of the interaction pathways that enables us to qualitatively explain features observed in the polariton spectra and dynamics of observables obtained via numerical simulations. A central finding of our study is that the angular momentum transfer pathways are dramatically altered when going from one~\cite{nandipati2023cavity} to even just two molecules. We note that JT Hamiltonians are a cornerstone for the study of molecular spectroscopy and dynamics~\cite{bersuker2006jahn, koppel} and have also recently become the focus of analogue quantum simulation experiments~\cite{valahu2023direct,whitlow2023quantum,wang2023observation}.

\emph{Model.---} We consider $N$ JT active molecules coupled identically to two orthogonally polarized modes of a FP  cavity, as depicted in Fig.~\ref{fig:setup}. Each molecule $k$ is described by two degenerate excited electronic levels $\ket{E_x}$ and $\ket{E_y}$ that are well separated by an energy $\epsilon =\hbar\omega_{\epsilon}$ from the electronic ground state $\ket{A}$ at a reference nuclear geometry. By the symmetry selection rules for JT vibronic coupling~\cite{bersuker2006jahn}, the $\ket{E}$ states couple via the degenerate $e$ vibrational modes $Q_x$ and $Q_y$ of the molecule, each with frequency $\omega$ and described respectively by ladder operators $\hat{b}_x^{(\dagger)}$ and $\hat{b}_y^{(\dagger)}$. 
%which is termed as the $\Exe$ JT interaction~{\color{red} ref}. 
In terms of the linear combinations $\ket{E_\pm} = (\ket{E_x}\pm i\ket{E_y})/\sqrt{2}$ and $\hat{b}_\pm ^{(\dagger)}= (\hat{b}_x \pm i\hat{b}_y) ^{(\dagger)}/\sqrt{2}$
the resulting Hamiltonian for molecule $k$ is given by~\cite{longuet1958studies,nandipati2023cavity} ($\hbar=1$)
\begin{align}\label{eqn:ham_jt}
    \hat{H}_{m,k} & = \sum_{r=+,-} \left[ \omega \hat{b}_{r,k}^\dagger \hat{b}_{r,k}  + \omega_{\epsilon}  \ket{E_r}_k\bra{E_r}_k  \right] \\\nonumber
     & + \kappa \left[ \left(\hat{b}_{+,k}^\dagger + \hat{b}_{-,k} \right)\ket{E_-}_k\bra{E_+}_k + \text{h.c.} \right],
\end{align}
The second line of Eq.~(\ref{eqn:ham_jt}) describes the linear $\Exe$ JT interaction with strength $\kappa$.

Introducing the cavity, the $\ket{A}\leftrightarrow\ket{E_\pm}$ molecular excitations are coupled to two degenerate, circularly polarized cavity modes with frequency $\omega_c$ [as shown in Fig.~\ref{fig:setup}(b)], that are described by the operators $\hat{a}_\pm^{(\dagger)}$~\cite{cri_91_2430}. Within the rotating wave approximation, the total Hamiltonian for the cavity-molecular system is given by 
\begin{equation}
    \hat{H} =  \hat{H}_c + \sum_{k=1}^N \left(\hat{H}_{m,k}+\hat{H}_{m-c,k}\right),
    \label{eqn:ham_tot}
\end{equation}
where $\hat{H}_c = \omega_c\left(\hat{a}_+^\dag\hat{a}_+ + \hat{a}_-^\dag\hat{a}_-\right)$ is the free cavity Hamiltonian, $\hat{H}_{m,k}$ is the molecular Hamiltonian, Eq.~(\ref{eqn:ham_jt}), and
\begin{equation}
    \hat{H}_{m-c,k} = \frac{\Omega}{2\sqrt{N}}\left(\hat{a}_+^\dag\ket{A}_k\bra{E_+}_k + \hat{a}_-^\dag\ket{A}_k\bra{E_-}_k + \;{\rm{h.c.}}\right)
    \label{eqn:ham_mc}
\end{equation}
describes the coupling of molecule $k$ to the cavity with strength $\Omega/(2\sqrt{N})$. The normalization by $\sqrt{N}$ compensates for the the collective enhancement in molecule-cavity coupling strength so that it becomes approximately independent of $N$. 

\begin{figure}[t]
  \centering
  \includegraphics[width=0.8\columnwidth]{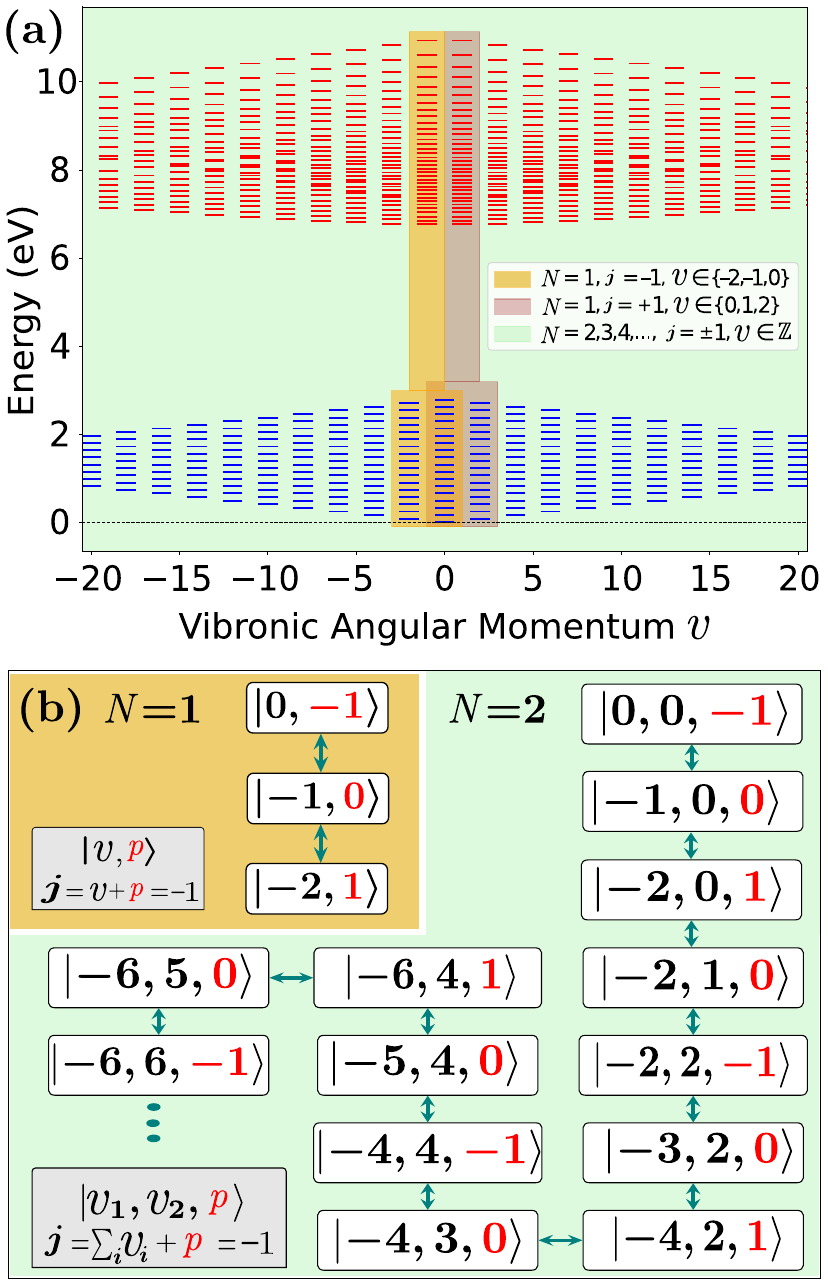}
  \caption{(a) Vibronic spectrum of a single JT molecule. The energy levels are grouped into different vibronic angular momentum sectors, labeled by $v$. (b) Collective vibronic angular momentum cascade: The mechanism of molecule-cavity coupling for the case of single JT molecule ($N$=1, shaded in orange) and for the case of two JT molecules ($N$=2, shaded in green) coupled to the cavity (cf. Fig.~\ref{fig:setup}). In the latter case, starting from a single right-circularly polarized cavity photon and the two molecules in the ground state, the vibronic-photonic coupling leads to a proliferation of $v$ values for the two molecules. In contrast, the maximum value of accessible $v$ is restricted to $2$ in the $N$=1 case. The coupling mechanism follows the conservation of total vibronic-photonic angular momentum $j$ in both the cases.
  }
  \label{fig:vibronic_ang_mom}
\end{figure}

\emph{Vibronic angular momentum.---} The JT molecular Hamiltonian, Eq.~(\ref{eqn:ham_jt}), involves an exchange of angular momentum between the electronic and vibrational degrees of freedom \cite{Longuet-Higgins,nandipati2021dynamical}; defining an electronic angular momentum operator $\hat{S}_{z,k} = \ket{E_+}_k\bra{E_+}_k-\ket{E_-}_k\bra{E_-}_k$ and a vibrational angular momentum operator $\hat{L}_{z,k} = \hat{b}_+^\dag\hat{b}_+ -  \hat{b}_-^\dag\hat{b}_-$, we observe that $\hat{H}_{m,k}$ commutes with the total \emph{vibronic} angular momentum operator~\footnote{This quantity is denoted as $2l$ in Ref.~\cite{Longuet-Higgins}.} 
\begin{equation}
    \hat{V}_k = 2\hat{L}_{z,k} + \hat{S}_{z,k}. 
    \label{eqn:vibr_ang}
\end{equation}
Hence, the eigenstates of $\hat{H}_{m,k}$ have well-defined vibronic quantum numbers $v$ and can be grouped into sectors of fixed $v$. In terms of these \emph{vibronic} eigenstates, $\hat{H}_{m,k}$ can be written in a diagonal form as 
\begin{equation}
    \hat{H}_{m,k} = \sum_{v}\sum_{i}\lambda_{i}^v\ket{\lambda_i^v}_k\bra{\lambda_i^v}_k,
    \label{ham_vib_diag}
\end{equation}
where $\lambda_i^v$ and $\ket{\lambda_i^v}$ denote the $i$th eigenfrequency and eigenstate in sector $v$. Figure~\ref{fig:vibronic_ang_mom}(a) shows the spectrum of $\hat{H}_{m,k}$, with the energy levels grouped into sectors of fixed $v$. The ground manifold consists of trivial eigenstates of the form $\ket{A,n_+,n_-}$, where $n_\pm$ denote Fock states of the vibrational  modes, and hence only populate sectors with even $v$ values, since $\hat{V}_k \ket{A,n_+,n_-} = 2(n_+-n_-)\ket{A,n_+,n_-}$. The excited manifold has non-trivial eigenstates formed by the JT interaction, and since $\hat{S}_{z}\ket{E_\pm}=(\pm 1)\ket{E_\pm}$, these states only populate odd $v$ sectors. 

\emph{Qualitative discussion of molecule-cavity interaction.---} The total Hamiltonian~(\ref{eqn:ham_tot}) conserves two quantum numbers, which in turn facilitate qualitative understanding of the molecule-cavity interaction; first, $\hat{H}$ commutes with the operator corresponding to the total number of electronic and photonic excitations $\hat{N}_{\rm ex}$, defined as
\begin{equation}
    \hat{N}_{\rm ex} = \sum_{k=1}^N \left(\ket{E_+}_k\bra{E_+}_k + \ket{E_-}_k\bra{E_-}_k \right) + \hat{a}_+^\dagger \hat{a}_+ + \hat{a}_-^\dagger \hat{a}_-.
\end{equation}
A second conserved quantity is the total angular momentum. Introducing the photonic angular momentum operator $\hat{P} = \hat{a}_+^\dag\hat{a}_+ - \hat{a}_-^\dag\hat{a}_-$, we observe that $\hat{H}$ commutes with the total angular momentum operator 
\begin{equation}
    \hat{J} = \sum_{k=1}^N \hat{V}_k + \hat{P}.
\end{equation}
These two conservation laws arise respectively from the Jaynes-Cummings type nature of the cavity-molecule interaction and the dipole selection rules that enforce coupling of matching cavity polarization components to the corresponding vibronic transitions in each molecule. Denoting the quantum numbers associated with $\hat{N}_{\rm ex}$ and $\hat{J}$  by $n_{\rm ex}$ and $j$, the total cavity-molecular Hamiltonian~(\ref{eqn:ham_tot}) can thus be block-diagonalized into sectors of fixed $(n_{\rm ex},j)$.

Focusing on the $(n_{\rm ex}=1,j=-1)$ sector, we first consider the situation of a single JT-active molecule coupled to the cavity. In this case, the maximum cavity occupation is one photon in either the right or left circularly polarized (RCP/LCP) mode, resulting in the possible photonic angular momentum values of $p=-1$ (RCP), $p=1$ (LCP) or $p=0$ (no photon). Since the total angular momentum $j=-1$ is conserved, the Hamiltonian for the single molecule-cavity system can thus only populate molecular vibronic states with $v=0,-1,-2$. This is illustrated in Fig.~\ref{fig:vibronic_ang_mom}(b) using the basis $\ket{v,p}$, where we follow the states populated by the molecule-cavity interaction starting from a molecule in the vibronic sector $v=0$ and with one RCP photon ($p=-1$). For this exercise, we denote the molecular basis state solely by its vibronic quantum number $v$, although there are multiple vibronic eigenstates in each sector, as seen from Eq.~(\ref{ham_vib_diag}) and Fig.~\ref{fig:vibronic_ang_mom}(a). For the single-molecule case, we see that $\hat{H}$ can only redistribute population between three vibronic angular momentum sectors, and hence the range of $v$ is bounded. 

The situation is significantly different when we move to the case of two molecules coupled to the cavity. In the $(1,-1)$ sector, although the total vibronic momentum $v=v_1 + v_2$ of the two molecules is still limited to $0,-1,-2$, there is no restriction on the vibronic angular momenta $v_1$ and $v_2$ of the individual molecules. In Fig.~\ref{fig:vibronic_ang_mom}(b), we show one sequence of states populated by the Hamiltonian starting from both molecules in the $v_1=v_2=0$ sector and a single RCP photon. Similar to the single molecule case, we use the notation $\ket{v_1,v_2,p}$ to denote the basis states in order to track the sectors being populated. Unlike in the $N=1$ case, here the molecules can access states with high vibronic angular momentum, a phenomenon we term as collective vibronic cascade. The range of possible $v_k$ values accessed by the individual molecules depends upon the cavity frequency $\omega_c$ and the strength of the molecule-cavity coupling $\Omega$.

\begin{figure}[t]
  \centering
    \includegraphics[width=0.9\columnwidth]{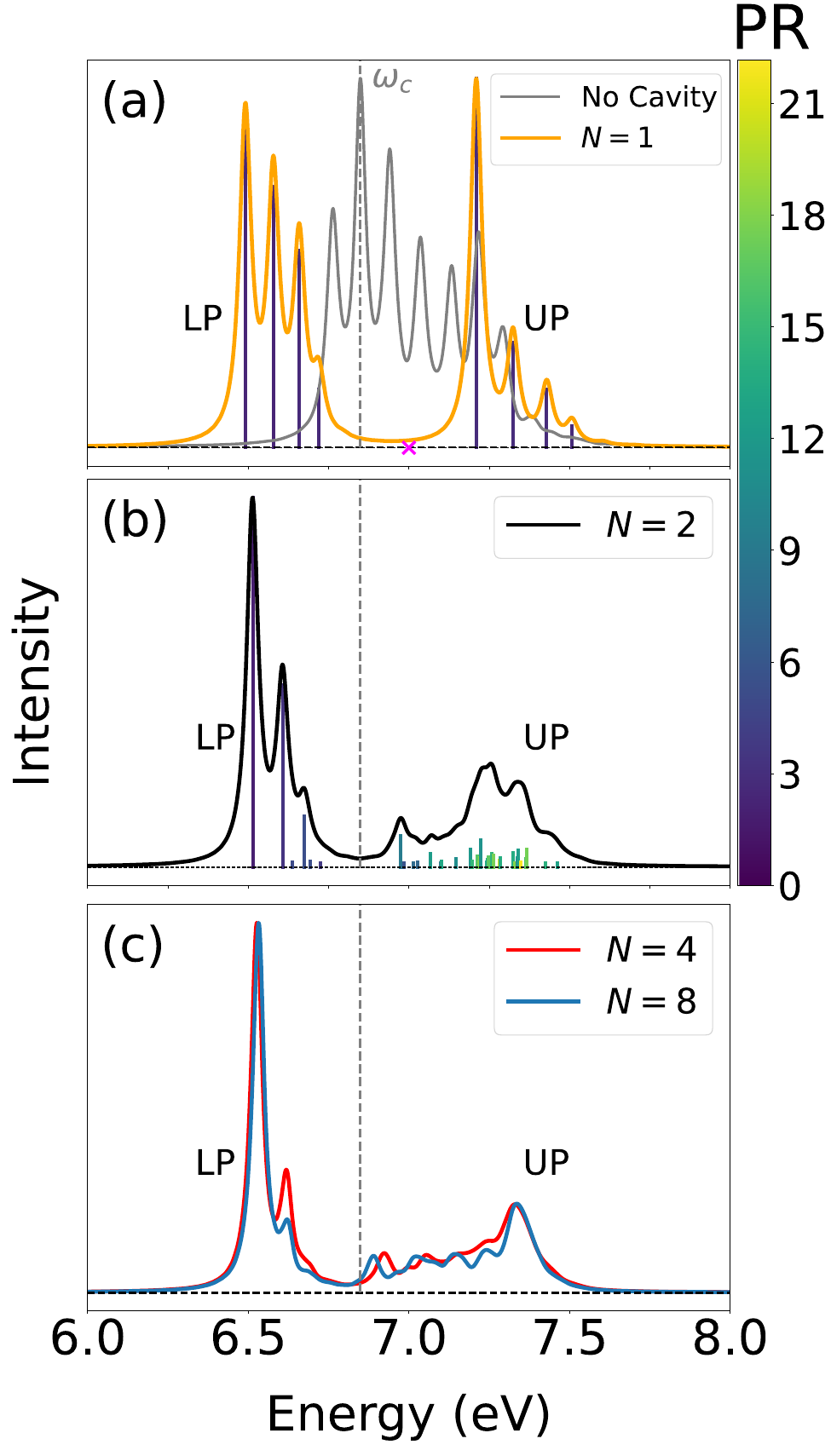}
  \caption{Polaritonic spectra for (a) $N$=1, (b)  $N$=2, and (c) $N$= 4 and 8 JT molecules coupled to the FP cavity. The spectra are obtained by setting the cavity resonant with the most optically bright JT vibronic state (shown in (a), grey); the '$\textcolor{magenta}{\times}$' in (a) indicates the $\ket{A}\leftrightarrow\ket{E_\pm}$ energy separation $\epsilon$.  The stick spectra in (a, b) are obtained by the diagonalization of the total Hamiltonian in Eq.~(\ref{eqn:ham_tot}). The color of the sticks indicates the participation ratio of the vibronic sectors, which quantifies the range of accessible vibronic angular momenta $v$ for each JT molecule. The low-resolution spectra (solid curves) in all the panels are obtained using the MCTDH method~\cite{mctdh_2}.}
  \label{fig:abs_spectra}
\end{figure}

\emph{Polariton spectra.---} We now quantitatively study the consequences of the collective vibronic cascade on the polaritonic spectrum, which is the absorption spectrum obtained by initializing the molecule-cavity system with one RCP photon in the cavity and the molecule(s) in their vibronic ground state.
We assume the $\ket{A}\leftrightarrow\ket{E_\pm}$ energy splitting  $\epsilon = 7$ eV, a vibrational frequency $\omega = 0.08196$ eV and a vibronic coupling strength $\kappa/\omega =  2.2$, which are typical for the excited states of the aromatic molecules, for example, sym-triazine~\cite{whetten1986dynamic}. Figure~\ref{fig:abs_spectra} summarizes the polaritonic spectra obtained numerically for $N=1,2,4$ and $8$ molecules coupled to a cavity. For reference, Fig.~\ref{fig:abs_spectra}(a) also shows the absorption spectrum of a single molecule in the absence of a cavity~\cite{nandipati2023cavity}. All the polaritonic spectra are computed by setting $\omega_c \approx 6.85$ eV, which makes the cavity resonant with the transition from $\ket{A,0,0}$ to the most optically bright eigenstate of the JT spectrum in the absence of the cavity, as indicated in Fig.~\ref{fig:abs_spectra}(a). We set the cavity-molecule coupling strength to be $\Omega/(2\omega_c)=0.05$.

For $N\leq 2$, we utilize the symmetries presented by the Hamiltonian and numerically diagonalize it to obtain the exact eigenstates of the system. We focus on the eigenstates in the $(n_{\rm ex}=1, j=-1)$ sector, whose energies are given by the locations of the discrete stems in Fig.~\ref{fig:abs_spectra}(a) and~(b). We consider probing the system by weak cavity driving, and hence the absorption intensities of the polaritonic states are obtained as their overlap with the basis state consisting of all molecules in the vibronic ground state and a single photon in the RCP mode.

Complementing the diagonalization procedure, we have performed quantum dynamical calculations for obtaining low-resolution spectra, shown as continuous solid lines in Fig.~\ref{fig:abs_spectra}, using the Multi-Configuration Time-Dependent Hartree (MCTDH)  method~\cite{mctdh_2,vendrell2018coherent}. After benchmarking this method with the exact diagonalization approach for $N=1$ and $2$ molecules, we use it to extend our numerical studies to larger systems of $N=4$ and $8$ molecules, where the latter is computationally prohibitive. Further details of the numerical methods are discussed in the Supplemental Material (SM)~\cite{SI}.

In all the cases shown in Fig.~\ref{fig:abs_spectra}, the polaritonic spectrum splits into a lower (LP) and upper polariton (UP) branch. For $N=1$, both branches consist of well-resolved peaks, consistent with $4$ eigenstates in each branch~\cite{nandipati2023cavity}. In comparison, for $N=2$, while the LP branch is still fairly well-resolved, the UP branch is significantly broadened. This feature is a direct consequence of the collective vibronic cascade effect: As the number of accessible vibronic sectors is in principle unbounded, a large number of polariton states are formed that respond weakly to cavity driving, as evidenced by the numerous short stems in the UP branch in Fig.~\ref{fig:abs_spectra}(b). As a measure of the number of participating vibronic sectors, we compute the sector participation ratio, ${\rm PR} = 1/\left(\sum_v P^2(v)\right)$. Here, $P(v)$ is the probability for a molecule to occupy sector $v$, computed by summing the probabilities to occupy all the vibronic eigenstates $\ket{\lambda_i^v}$ belonging to that sector~\cite{SI}. The color of the stems shown in Fig.~\ref{fig:abs_spectra}(a) and~(b) indicate the PR of the corresponding polaritonic eigenstate. For $N=1$, since only $3$ vibronic sectors can be populated, PR $\leq 3$, as seen from the stem colors in Fig.~\ref{fig:abs_spectra}(a). On the other hand, for $N=2$, the states in the UP branch are found to have PR values as high as 20, indicating that a large number of vibronic sectors are accessed by the molecules when they form these polariton states. In the SM, we plot a heatmap showing the population distribution across vibronic sectors for the $N=2$ case, which further demonstrates the vibronic cascade effect~\cite{SI}. Increasing the number of molecules to $N=4$ and $8$ [Fig.~\ref{fig:abs_spectra}(c)], the polariton spectrum is noticeably broadened in both the LP and UP branches, which can be intuitively understood to be a consequence of larger number of available pathways for the cavity-mediated redistribution of vibrational angular momentum between molecules.

\begin{figure}[t]
  \centering
 \includegraphics[width=0.95\columnwidth]{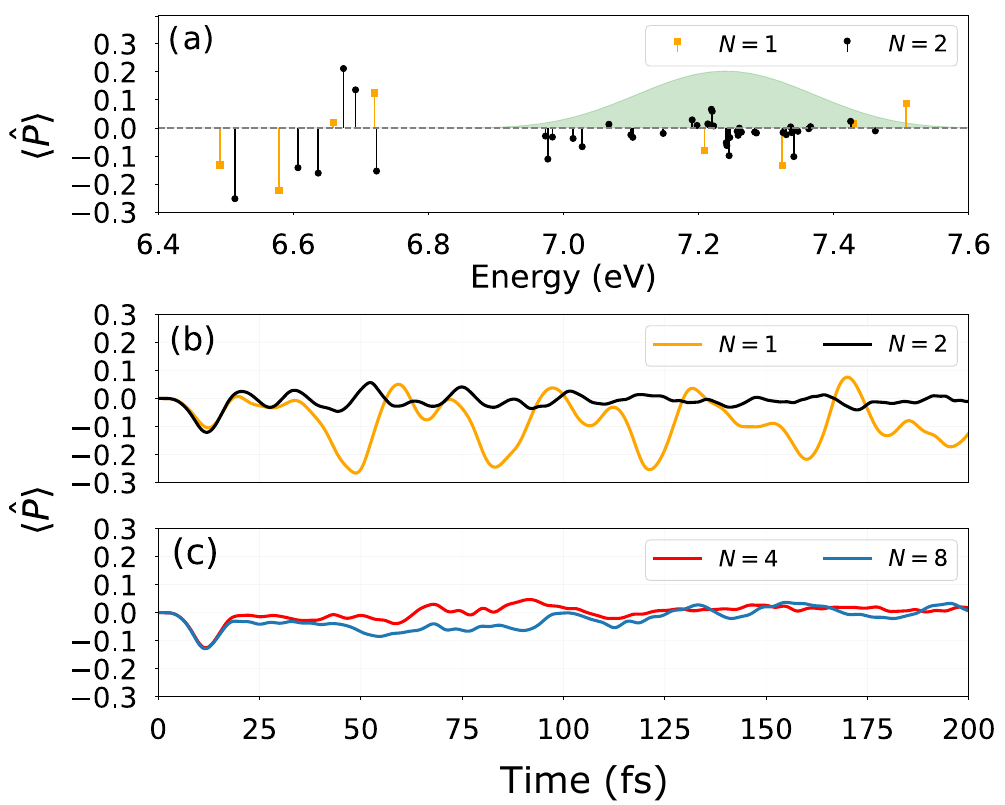}
 \caption{Cavity-polarization dynamics under broadband excitation. (a) Net cavity-polarizations associated with polaritonic states belonging to the $(1, -1)$ sector for $N$ = 1 (orange) and $N=$ 2 (black). (b) Time-dependent response of cavity-polarizations to an external RCP pulse for (b) $N$=1 and 2, and (c) $N$= 4 and 8. The polarizations are obtained by targeting the UP branch with a short $20$ fs pulse of bandwidth $\sim0.25$ eV (shaded in green, in (a)). Exactly the same time-dependent response, but with an opposite sign, can be triggered with an LCP pulse by targeting the $(1,1)$ sector.}
\label{fig:cav_polarization} 
\end{figure}

\emph{Photon polarization and dynamics.---} An observable consequence of the collective vibronic cascade manifests in the net photon polarization $\ev{\hat{P}}$. Figure~\ref{fig:cav_polarization}(a) shows $\ev{\hat{P}}$ of the polaritonic eigenstates in the $(1,-1)$ sector, for the $N=1$ (orange) and $N=2$ (black) cases. This quantity is positive or negative depending on the relative  contributions of the LCP and RCP photons in the polaritonic state, with $-1\leq \ev{\hat{P}}\leq 1$. While $\ev{\hat{P}}$ is in general an order of magnitude less than $1$ in both the LP and UP branches for $N=1$, it is noticeably suppressed further for the majority of the states in the UP branch for $N=2$. This suppression can be qualitatively understood as resulting from the repeated mixing of LCP and RCP photons in the polariton formation because of the larger number of accessible vibronic sectors.
To observe the effect of the collective vibronic cascade on the cavity polarization dynamics, we target the states in the UP branch with a broadband RCP excitation pulse as illustrated in Fig.~\ref{fig:cav_polarization}(a). The pulse definition and parameters are provided in the SM~\cite{SI}. Starting from the molecule-cavity ground state, application of a weak, short RCP pulse prepares a coherent superposition of the polaritonic eigenstates in the $(1,-1)$ sector that fall within the width of the pulse spectrum. In Fig.~\ref{fig:cav_polarization}(b) and ~(c), we plot the resulting time-dependent polarization $\ev{\hat{P}(t)}$ for $N=1,2,4$ and~$8$, after normalizing it with respect to the total excitation probability. The normalization ensures that $\ev{\hat{P}(t)}$ is intensity independent for perturbative driving~\cite{nandipati2023cavity,nandipati2021dynamical}. Remarkably, as seen in Fig.~\ref{fig:cav_polarization}(b), the  polarization dynamics is significantly suppressed for the case of $N=2$ (black) as compared to the $N=1$ case (orange) where the magnitude of oscillations is significant. The trend of suppressed oscillations further continues in the case of $N=4$ and $N=8$, as shown in Fig.~\ref{fig:cav_polarization}(c), leading to very small values of the time-averaged polarization. The quenching of the net polarization is a direct consequence of the collective vibronic cascade as the increasing number of molecules offer myriad vibronic angular momentum redistribution pathways that scramble the photonic angular momentum.

\emph{Conclusion and Outlook.--} In conclusion, we have observed that JT molecules collectively coupled to a common cavity can access a cascade of high-angular-momentum vibronic states in the presence of a single cavity photon, in sharp contrast to the single molecule case where the range of accessible vibronic sectors are bounded. As a consequence of this collective vibronic cascade, the resulting polaritonic spectra are broadened, and the net polarization of individual eigenstates and the polarization dynamics subsequent to weak excitation is strongly suppressed. In the future, our analysis can be extended to very large numbers of JT molecules in cavities, which is the typical situation in current experiments, using mean-field theory and related semiclassical methods that become accurate with increasing $N$~\cite{thanh2025semiclassical,Reitz,fregoni2022theoretical}. Such methods could also be used to go beyond the harmonic approximation for the molecular vibrations~\cite{mondal2024cavity} in order to study the implications of the collective effects demonstrated here on chemical reaction pathways. The inclusion of disorder in molecular energies and orientations, along with dissipation of the cavity mode~\cite{wellnitz2022disorder,bond2024open}, will make this system an even richer platform to study many-body physics and chemistry. Finally, going to the regime of strong light-matter interactions~\cite{mandal2023theoretical} will cause a breakdown of the rotating-wave approximation and various conservation laws, and can hence unlock new collective phenomena beyond what we have reported here.\vspace{1em}

 \begin{acknowledgments}
 We thank Prof. Wolfgang Domcke and Prof. Srihari Keshavamurthy for useful comments on the manuscript, and Jayden Koshy Joe for his help with numerics.  K.R.N. thanks Prof. Oriol Vendrell for preliminary discussions. A.P. and S.K.P. acknowledge financial support from IIT Madras. K.R.N. thanks IIT Madras for the financial and infrastructural support through the start-up grant (NFSC). A.S. acknowledges support by the Department of Science and Technology, Govt. of India through the INSPIRE Faculty Award, by the Anusandhan National Research Foundation (ANRF), Govt. of India through the Prime Minister's Early Career Research Grant (PMECRG), and by IIT Madras through the New Faculty Initiation Grant (NFIG). 
\end{acknowledgments}

%\bibliography{JT-cavity.bib}
%apsrev4-2.bst 2019-01-14 (MD) hand-edited version of apsrev4-1.bst
%Control: key (0)
%Control: author (8) initials jnrlst
%Control: editor formatted (1) identically to author
%Control: production of article title (0) allowed
%Control: page (0) single
%Control: year (1) truncated
%Control: production of eprint (0) enabled
%

\end{document}

\newcommand{\affA}{Department of Chemistry, Indian Institute of Technology Madras, 600036 Chennai, India}
\newcommand{\affB}{Department of Physics, Indian Institute of Technology Madras, 600036 Chennai, India}
\newcommand{\affC}{Center for Quantum Information, Communication and Computing, Indian Institute of Technology Madras, Chennai 600036, India}

\affiliation{\affA}
\affiliation{\affB}
\affiliation{\affC}

\author{Suraj Kumar Pandit}
\thanks{These authors contributed equally to this work.}
\affiliation{\affA}\affiliation{\affC}

\author{Abhinay Pandey}
\thanks{These authors contributed equally to this work.}
\affiliation{\affB}\affiliation{\affC}

\author{Athreya Shankar}
\email{athreya@iitm.ac.in}
\affiliation{\affB}\affiliation{\affC}
\thanks{Corresponding author}

\author{Krishna R. Nandipati}
\email{knandipati@iitm.ac.in}
\affiliation{\affA}\affiliation{\affC}
\thanks{Corresponding author}

\title{Supplementary Material:\\
Collective Vibronic Cascade in Cavity-Coupled Jahn-Teller Active Molecules}

\date{\today}

\maketitle

\renewcommand{\theequation}{S\arabic{equation}}
\renewcommand{\thefigure}{S\arabic{figure}}
\renewcommand{\thetable}{S\arabic{table}}

\section{Diagonalization of Cavity-JT Molecular Hamiltonian}

For the numerical diagonalization of the cavity-molecule system, we first diagonalize the molecular Hamiltonian and obtain the vibronic eigenstates. Subsequently, we express the cavity-molecule Hamiltonian in the basis of vibronic eigenstates. The symmetry of the Hamiltonian under the exchange of molecular indices enables us to switch to an occupation number representation, instead of tracking the individual molecular states. Furthermore, the Hamiltonian conserves the total number of electronic and photonic excitations and the total angular momentum, defined as in Eqs.~(6) and~(7) of the main text. Hence, we can further reduce the Hamiltonian dimension greatly by directly constructing it in the subspace of basis states that satisfy $j=-1$ and $n_{\rm ex} = 1$, which is the sector of interest in this work. Further details of the vibronic eigenstates and the occupation number representation are provided in the subsections below.

\subsection{Vibronic basis}

Diagonalizing Equation~(1) in the main text, we obtain the vibronic eigenvectors of $\hat{H}_{m,k}$ 

\[
\ket{\lambda_j^v}_k = \sum_{s,n_+,n_-}C_{n_+n_-}^s\ket{s,n_+,n_-}_k
\]
with $s\in \{A,E_+,E_-\}$. The vibronic eigenvectors can be classified as trivial and non-trivial eigenvectors based on the absence or presence of vibronic coupling. The trivial eigenvectors, marked in blue in Fig.~2(a), are of the form $\ket{\lambda_j^v} = \ket{A,n_+,n_-}$ and belong to even $v$ sectors. The non-trivial eigenvectors, marked in red in Fig.~2(b), have $C^{A}_{n_+n_-} = 0$ and are superpositions of basis states of the form $\ket{E_\pm,n_+,n_-}$ of fixed $v$. These belong to odd $v$ sectors.

We use two indices to label the vibronic eigenvectors, the vibronic sector $v$ and the index $j$ that labels eigenvectors within each sector in order of increasing energy. For compactness, we use the symbol $i\equiv (v,j)$ to label vibronic eigenvectors in the following. We can write the full molecule-cavity Hamiltonian, Equation~(2), in the vibronic basis via a unitary transformation with the matrix 

\[
\hat U = \bigotimes_{k=1}^N \hat{U}_k\quad \text{, where} \quad \hat{U}_k = \sum_{s,n_+,n_-,i}U_{sn_+n_-,i}\ket{sn_+n_-}\bra{\lambda_i},
\]
using which $\hat{H}'=\hat{U}^\dagger\hat{H}\hat{U}$ is given by 
\begin{equation}
\label{eqn:ham_cjt_vib_basis}
\hat{H}' = \sum_{i,k} \lambda_i \ket{\lambda_i}_k\bra{\lambda_i}_k 
+ \omega_c \big(\hat{a}_+^\dagger \hat{a}_+ + \hat{a}_-^\dagger \hat{a}_-\big)
\;\; + \frac{\Omega}{2\sqrt{N}}\sum_{i,j,k}\Big(
 \alpha_{i,j}\, \ket{\lambda_i}_k\bra{\lambda_j}_k\hat a^\dagger_+ + \beta_{i,j}\, \ket{\lambda_i}_k\bra{\lambda_j}_k\hat a_-^\dagger + h.c 
\Big)
\end{equation}

where, $\alpha_{i,j} = \sum\limits_{n_+,n_-} U^\dagger_{A, n_+, n_-, i}\, U_{E_+, n_+, n_-, j}$ and $\beta_{i,j}   = \sum\limits_{n_+,n_-} U^\dagger_{A, n_+, n_-, i}\, U_{E_-, n_+, n_-, j}$. Thus, in the vibronic basis, the molecule-cavity interaction can be interpreted as a multi-level, two-mode generalized Tavis-Cummings model with the cavity modes mediating transitions between various pairs of levels in the high-dimensional molecular system. 

\subsection{Occupation number representation}

We assume that all the molecules are coupled identically to both the cavity modes. Hence, the Hamiltonian~~(\ref{eqn:ham_cjt_vib_basis}) is invariant under a permutation of molecule indices $k$. As a result, we can efficiently analyze the eigenstates and dynamics within the particle-permutation symmetric subspace by switching to an occupation number representation. That is, instead of tracking the states of individual molecules, we describe the collective state of the molecules by specifying the number of molecules in each vibronic state. 

For a system of $N$ molecules having $d$ levels, the particle-permutation symmetric basis states can be expressed in terms of the individual molecular states as 
\begin{align}
|N_0, N_1, \ldots N_{d-1} \rangle = \frac{1}{\sqrt{\frac{N!}{\prod\limits_{i=0}^{d-1} N_i!}}} P_{d}(\ket{\underbrace{0,0,0\dots}_{N_0 },\underbrace{1,1,1\dots}_{N_1 },\dots,\underbrace{d-1,d-1,d-1\dots}_{N_{d-1} } }).
\end{align}
Here, the symbol $P_d$ denotes the operation of summing over all possible arrangements of $N_0,N_1,\ldots,N_{d-1}$ labels of type $0,1,\ldots,d-1$ respectively, among the available $N$ molecules, with $\sum_i N_i = N$. The normalization constant is just the square root of the multinomial coefficient that gives the number of possible arrangements.

The Hamiltonian~(\ref{eqn:ham_cjt_vib_basis}) can be expressed using collective molecular operators that act on the occupation number basis states. First, the self-energy terms of the molecules can be written in terms of diagonal operators $\hat D_l =\sum\limits_{l=1}^{N} \ket{\lambda_i}_l\bra{\lambda_i}_l$, which have the property 
\begin{equation}
\label{eqn:diagonal_operator}
    \hat{D}_l\ket{N_0, N_1,..., N_l,..,  N_{d-1}} = N_l\ket{N_0, N_1,..., N_l,..,  N_{d-1}}.
\end{equation}
Next, the molecule-cavity interaction can be expressed in terms of collective ladder operators $\hat{T}^-_{j,k} = \sum\limits_{l=1}^{N} \ket{\lambda_j}_l\bra{\lambda_k}_l$ that remove one molecule from $\ket{k}$ and add it to $\ket{j}$, i.e.
\begin{equation}
\label{eqn:ladder_operator}
    \hat{T}^-_{j,k} \ket{N_0\dots N_j,\dots, N_k\dots N_{d-1}} = \sqrt{N_k(N_j+1)}\ket{N_0 \ldots N_j+1 \dots N_k-1 \dots  N_{d-1} }.
\end{equation}
Analogously, we can describe the conjugate process using 
\begin{equation}
\label{eqn:ladder_operators}
    \hat{T}^+_{j,k} \ket{N_0\dots N_j,\dots, N_k\dots N_{d-1}} = \sqrt{N_j(N_k+1)}\ket{N_0 \ldots N_j-1 \dots N_k+1 \dots  N_{d-1} }.
\end{equation}

Using these collective operators, the Hamiltonian~(\ref{eqn:ham_cjt_vib_basis}) can be expressed as 
\begin{equation}
\label{eqn:Ham_in_vib_basis}
    \hat H' = \sum_i \lambda_i \hat{D}_i  + \omega_c (\hat a_+^\dagger \hat a_+ +\hat a_-^\dagger \hat a_-) + \frac{\Omega}{2\sqrt{N}} \sum_{i.j}(\alpha_{i,j}\hat{T}^-_{i,j}\hat a^\dagger_+ +\beta_{i,j}\hat{T}^-_{i,j}\hat a^\dagger_- + {\rm h.c.} ).
\end{equation}

In practice, we diagonalize the molecular Hamiltonian using a truncated Fock space of $18$ Fock states for each vibrational mode. Subsequently, we construct the cavity-molecule Hamiltonian in the symmetric subspace of $N$ molecules by using the occupation number representation and associated collective molecular operators. For the $N=1$ and $N=2$ cases, the dimensions of the Hamiltonian obtained using this approach are $70\times 70$ and $11664\times 11664$ respectively. In contrast, a naive construction of the $N=2$ molecule-cavity Hamiltonian without leveraging any of the symmetries would lead to a Hamiltonian of dimension $3779136\times3779136$ ($3$ electronic levels and $18$ Fock bases for each of the $2$ vibrational modes of each molecule, and $2$ cavity modes with $2$ Fock bases each).

\section{Participation Ratio}

In this section, we describe the procedure to extract the populations in the vibronic sectors for computing their participation ratio (PR) in the polaritonic eigenstates. 

For $N = 1$, the polaritonic eigenstates are of the form 
\begin{equation}
    \ket{\psi} = \sum_{v,i,p}C_{i,p}^v \ket{\lambda_i^v,p},
\end{equation}
where $\ket{\lambda_i^v}$ denotes the $i$th vibronic eigenstate in sector $v$, and $p=0,\pm 1$ indicates no cavity photons or one photon in the LCP or RCP modes respectively. Hence, the total occupation of a vibronic sector $v_0$ is given by 
\begin{equation}
    P(v) = \sum_{i,p} \abs{C_{i,p}^{v_0}}^2,
\end{equation}
using which the PR can be calculated.

In the case of $N=2$, extracting the probability of any single molecule occupying a particular vibronic sector is slightly more involved. In the vibronic basis, the polaritonic eigenstates have the form 
\begin{eqnarray}
    \ket{\psi} = \sum_{v,i,p}C_{i,p}^{v}\ket{\lambda_i^v,\lambda_i^v,p} + \sum_{\substack{v,i,j,p\\j>i}} C_{i,j,p}^{v}\frac{1}{\sqrt{2}}\left(\ket{\lambda_i^{v},\lambda_j^{v}} + \ket{\lambda_j^{v},\lambda_i^{v}} \right)\ket{p} \nonumber\\
    + \sum_{\substack{v_1,v_2,i,j,p\\v_2>v_1}} C_{i,j,p}^{v_1,v_2}\frac{1}{\sqrt{2}}\left(\ket{\lambda_i^{v_1},\lambda_j^{v_2}} + \ket{\lambda_j^{v_2},\lambda_i^{v_1}} \right)\ket{p}.
    \label{eqn:two_mol_cjt_state}
\end{eqnarray}
The first term represents terms where both molecules are in the same basis state. The second term corresponds to the situation where both molecules are in the same vibronic sector, but occupy different levels in that sector. Finally, the third term contains contributions arising from both molecules occupying different vibronic sectors. 

In this case, the probability to find, say molecule 1, in a sector $v_0$ is given by
\begin{eqnarray}
    P(v_0) = \sum_{i,p} \abs{C_{i,p}^{v_0}}^2 + \sum_{\substack{i,j,p\\j > i}} \abs{C_{i,j,p}^{v_0}}^2 + \frac{1}{2}\sum_{\substack{v,i,j,p\\v < v_0}}\abs{C_{i,j,p}^{v,v_0}}^2 + \frac{1}{2}\sum_{\substack{v,i,j,p\\v > v_0}}\abs{C_{i,j,p}^{v_0,v}}^2. 
\end{eqnarray}
Here, the first two contributions arise from the first two terms in Eq.~(\ref{eqn:two_mol_cjt_state}), while the third and fourth contributions arise from the last term. The factor of $1/2$ in these terms accounts for the symmetric structure of the two-molecule state, which gives a $1/2$ probability for the first molecule to be in sector $v$ or $v_0\neq v$.

\begin{figure}[tb]
    \centering
    \includegraphics[width=0.6\linewidth]{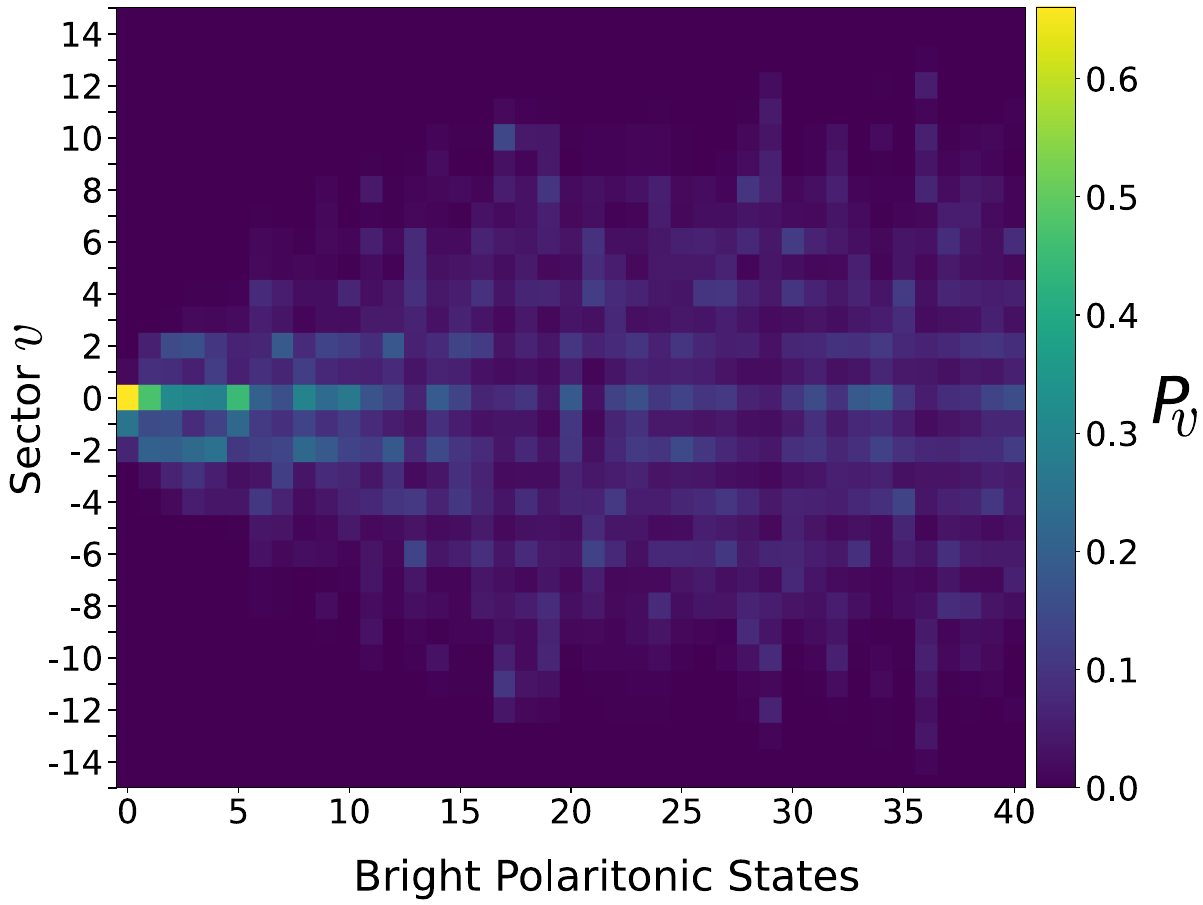}
    \caption{Heatmap showing the single-molecule occupation probability of different vibronic sectors $v$ for the various bright polaritonic eigenstates in the $N=2$ case.}
    \label{fig:heatmap}
\end{figure}

Figure~\ref{fig:heatmap} shows a heatmap of the occupation probabilities of different vibronic sectors for one of the two molecules in the $N=2$ case, when the molecule-cavity system is in the different bright polaritonic states. The polariton states are indexed in order of increasing energy and correspond to the discrete stems shown in Fig.~3(b) of the main text. The proliferation of vibronic angular momentum in the higher-lying polariton states is evident from this plot and explains the higher PR of these states, which is indicated by the colors of the stems in Fig.~3(b) of the main text. 

\newpage
\section{Dynamics  of Cavity-JT system with MCTDH method}

In the MCTDH approach~\cite{vendrell2018coherent,mctdh_2,mctdh}, the total wavefunction $\psi(\mathbf{q},t)$ is expanded as the Hartree-products of the so-called (time-dependent) single particle functions (SPFs) for the degrees of freedom $\mathbf{q} \equiv \{\mathbf{q}_v,\mathbf{q}_s, \mathbf{q}_p\}$, where $\mathbf{q}_v$, $\mathbf{q}_s$ and $\mathbf{q}_p$ are the vibrational, electronic  and photonic modes.  These SPFs are further expanded in a fixed number of (time-independent) primitive DVR basis functions for $\mathbf{q}_v$, $\mathbf{q}_s$ and $\mathbf{q}_p$.  In our dynamical calculations, we use the multi-layer (ML) version of the MCTDH wavefunction  $\psi(\mathbf{q}_v,\mathbf{q}_s,\mathbf{q}_p, t)$ ~\cite{vendrell2018coherent,larsson2024tensor} for treating the cavity-JT system dynamics (even under external laser pulse). The ML-form of $\psi$ reads as, 
\begin{equation}
 \psi (\mathbf{q}_{v},\mathbf{q}_{s}, \mathbf{q}_{p},t) = \sum_{j_{1} = 1}^{n_{1}} \sum_{j_{2}} ^ {n_{2}} \sum_{j_{3}} ^ {n_{3}}   A_{j_1,j_2,j_3} ^1 (t) \varphi_{j_1}^{(1;1)} (\mathbf{q}_p,t)   \varphi_{j_2}^{(1;2)} (\mathbf{q}_{v},t) \varphi_{j_3}^{(1;3)} (\mathbf{q}_{s}, t).
 \label{eqn:mlwavefun}   
\end{equation}

\begin{figure}[tb]
    \centering
    \includegraphics[width=0.55\linewidth]{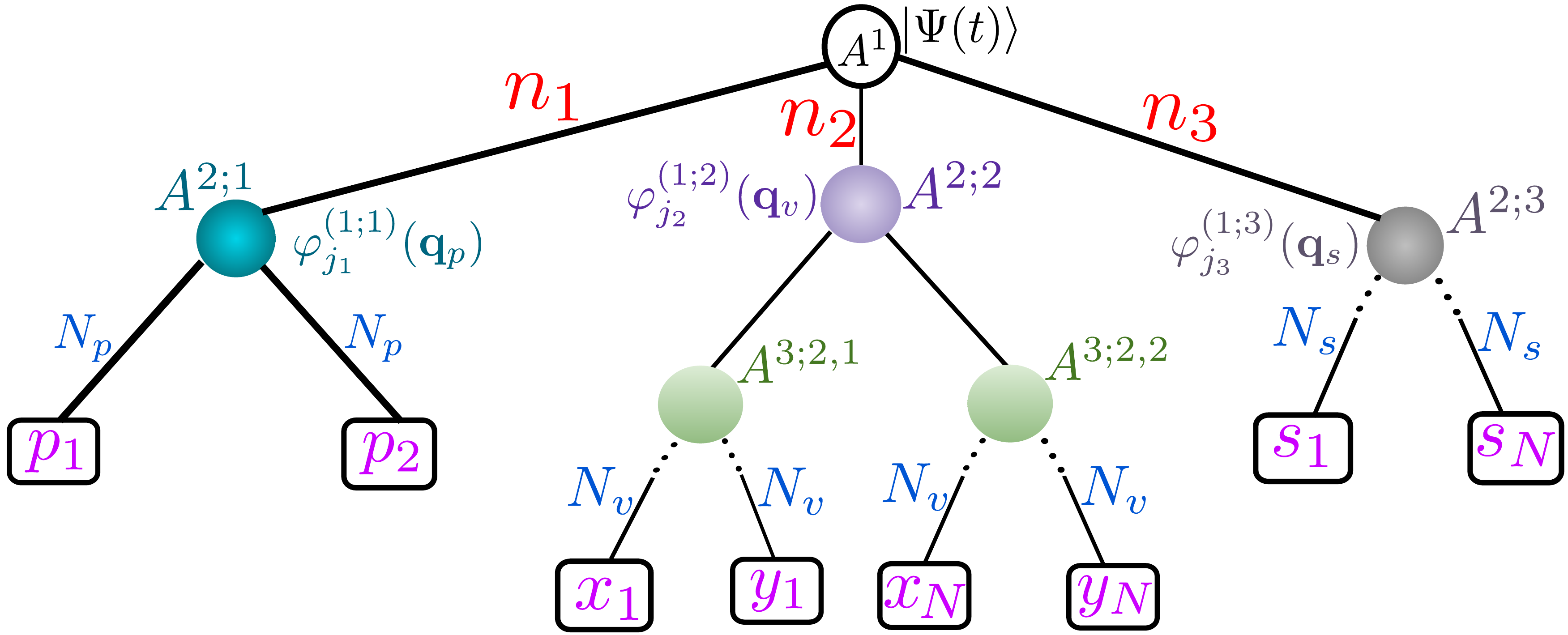}
    \caption{Multi-layer (ML) tree representation of the total wavefunction $\ket{\psi (t)}$ in Eq.~(\ref{eqn:mlwavefun}):In layer 1, the $\ket{\psi (t)}$ is expanded in the ($n_1$, $n_2$, $n_3$) SPFs of the photonic $\varphi_{j_1}^{(1;1)}$, vibrational $\varphi_{j_2}^{(1;2)}$ and electronic $\varphi_{j_3}^{(1;3)}$ degrees of freedom with A-coefficient at the top node. While the SPFs of photonic modes ($p_1, p_2$) are expanded as products of the primitive basis functions $N_p$ (in blue)  in the subsequent layer, the SPFs of the vibrational and electronic branches are further decomposed into MCTDH form in terms of lower dimensional SPFs in the layers below until the bottom layer is hit where a set of primitive basis functions $N_v$ and $N_s$ (in blue) are represented. The labels for the photonic ($p_1, p_2$), vibrational ($x_N, y_N$), and electroni ($s_N$) modes are given in pink at the bottom layer. We note that bare $N$ here runs over the number of molecules in the cavity, and not to be confused with primitive bases index $N_{p/v/s}$.}
    \label{fig:mltree}
\end{figure}

Here, $n_1,n_2$ and $n_3 $ respectively denote the number of SPFs for the photonic ($\varphi_{j_1}^{(1;1)}$), vibrational ($\varphi_{j_2}^{(1;2)}$) and electronic ($\varphi_{j_3}^{(1;3)}$) modes, and,  $A_{j_1,j_2,j_3} ^1 (t)$ is the corresponding time-dependent coefficient of the Hartree-products of these SPFs. The superscript ``1'' on the left of the semi-colon indicates layer number and the numbers (1, 2, 3) on the right represent the branch number in that layer. Pictorially, the assignment of the SPFs ($\varphi$) to the individual branches and the $A$-coefficient to a node connecting these branches is shown in  Fig~\ref{fig:mltree}. 

To begin with, the SPFs for the photonic mode, $\varphi_{i}^{(1;1)}$, can be directly expanded in terms of $\text{N}_p$ primitive basis functions $\chi_{m_1}(p_1)$ and $\chi_{m_2}(p_2)$ for the two photonic modes $p_1$ and $p_2$ through a linear combination of products of these functions (cf. the left side branch in Fig~\ref{fig:mltree}):

\begin{equation}
\varphi_{i}^{(1;1)} (\mathbf{q}_p,t) = \sum_{m_1,m_2=1} ^{\text{N}_p} A_{i,m} ^{2;1} (t) \chi_{m_1} (p_1) \chi_{m_2} (p_2)  \label{eqn:photonic}
\end{equation}
where $A_{j_1,i} ^{2;1}(t)$ are the time-dependent expansion coefficient of the subsequent layer containing the primitive bases. 

Next, we expand the SPFs for the vibrational mode, $\varphi_i^{(1;2)}$ using another layer of lower dimensional SPFs $\varphi_{j_{\lambda}}^{(2;2,\lambda)}$:
\begin{equation}
    \varphi_{i}^{(1;2)}(\mathbf{q}_{v},t) = \sum_{j_1=1} ^{n_{2,1}} \sum_{j_2=1} ^{n_{2,2}} \ldots \sum_{j_k =1} ^{n_{2,k}} A_{i;j_1,j_2,\dots,j_k} ^{2;2} (t) \prod _{\lambda=1} ^k \varphi_{j_{\lambda}} ^{(2;2,\lambda)} (\mathbf{q}_{v,\lambda} ,t).   \label{eqn:vib}
\end{equation}
Here, ${n_{2,1}}, n_{2,2}, \ldots, n_{2,k}$ denotes the number of SPFs for each vibrational sub-modes \({q}_{v,\lambda}\) of the branch $\lambda$ and $A_{i;j_1,j_2,\dots,j_k} ^{2;2} (t)$ is the usual time-dependent expansion coefficient in the layer 2 for branch 2, as shown in Fig.~\ref{fig:mltree}. 
Taking cavity-JT system with two  JT molecules (i.e. $N=2$) as an example, the SPFs in the Hartree-product on the right side for each sub-branch $\lambda$ can be expanded in the primitive basis for the vibrational modes like that of Eq.~(\ref{eqn:photonic}): 
\begin{equation}
    \varphi_i ^{(2;2,\lambda)} (\mathbf{q}_{v,\lambda} ,t) = \sum_{m_1,m_2=1} ^{\text{N}_v} A_{i,m} ^{3;2,\lambda} (t) \chi_{m_1}  (x_\lambda) \chi_{m_2}  (y_\lambda)  \label{eqn:vib_spf}
\end{equation}

In the case with more than 2 JT molecules in a cavity, i.e., $N > 2$,  the SPFs  $\varphi_{j_{2_,\lambda}} ^{(2;2,\lambda)}$ in Eq.~(\ref{eqn:vib}) are recursively expanded further in the MCTDH form in a set of SPFs of the lower-layer as in Eq.~(\ref{eqn:vib}) until the bottom layer is hit where the primitive basis for the modes is represented. This is shown with the dotted lines in the Fig.~\ref{fig:mltree} for the subsequent layers. Finally, the branch on the right side in Figure~\ref{fig:mltree}, representing electronic mode SPFs  $\varphi_{j_3}^{(1;3)} (\mathbf{q}_{s}, t)$ follow an analogous hierarchical decomposition to that of the vibrational branch.

\newpage
\begin{table}[tb]
\centering
    \caption{This Table presents the converged ML-tree (column 2) and the associated parameters for the dynamics of the cavity + $N$ JT molecules (column 1) system using the MCTDH method. The converged parameters of the dynamical calculations (including external CP pulse) are shown in the tree for each $N$ in Column 2: The number of SPFs  (in red) for each branch and the primitive basis functions (in blue) of the bottom layer in which photonic [($p_1$, $p_2$) on the left], vibration  [($x_m,y_m$) on the middle] and electronic degrees of freedom [($s_m$)] are given, where $m$ runs over the $N$ molecules. The primitive basis functions for the cavity and molecular vibrational modes were chosen to be sin-DVR and harmonic oscillator (HO-DVR) type, respectively, as defined in the MCTDH literature ~\cite{vendrell2018coherent,mctdh_2}. Here, we show the ML-tree for up to $N=4$ molecules, which by doubling the number of degrees of freedom, with the same number of SPFs and primitive functions, is found to give converged results for the $N=8$ case.}
    \includegraphics[width=0.6\linewidth]{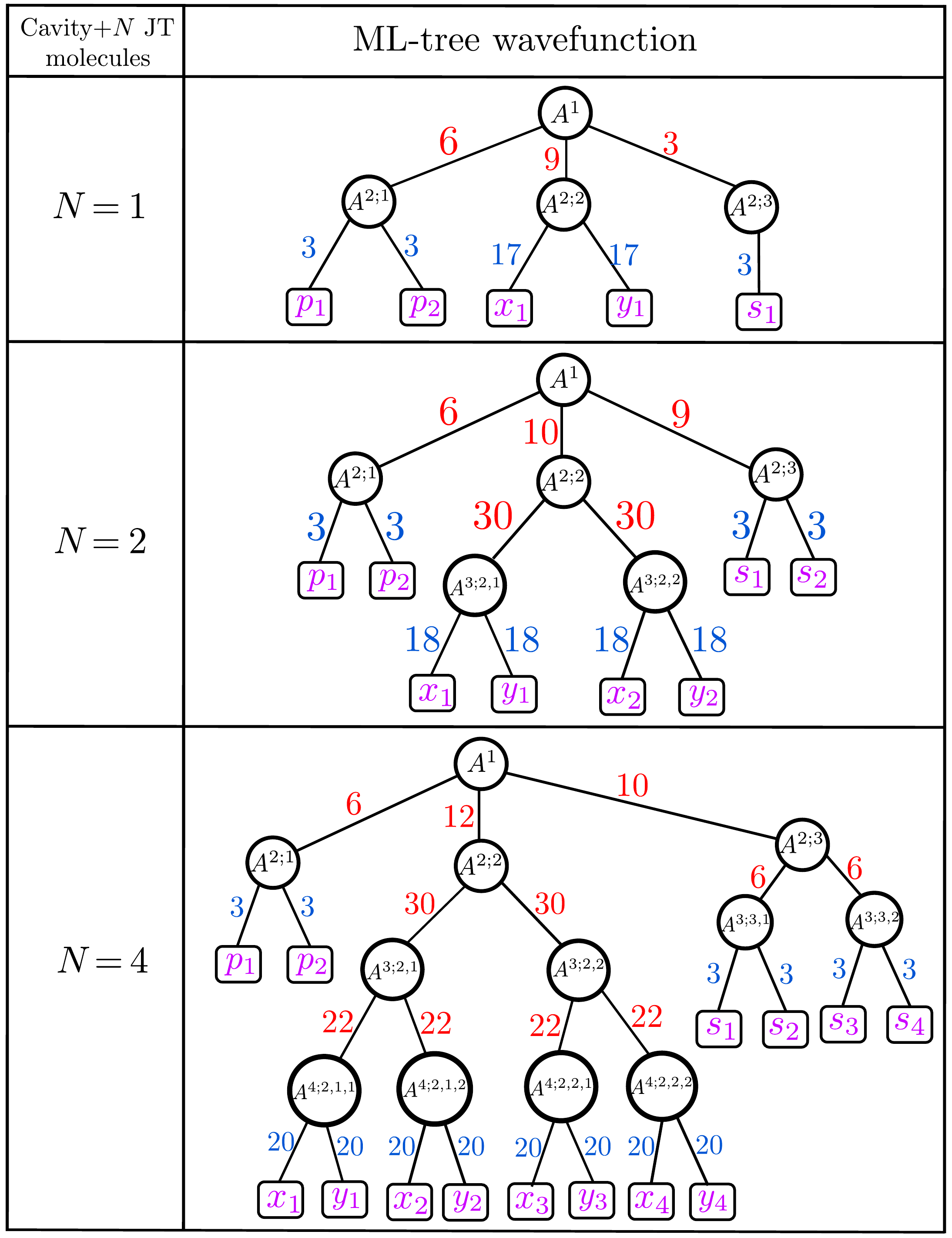} 
\end{table}
\newpage

\subsection{Dynamics under Circularly Polarized Light Excitation} 
The dynamics of cavity-molecule system under the excitation by a CP light is considered by driving the cavity subsystem.
As such, the interaction Hamiltonian between the dipole moment of the cavity subsystem and the external laser field is given by
\[ \hat{H}_{I} = - \hat{\vec{\mu}} . \vec{E}(t)  = - \left[ \hat{\mu}_x  E_x(t) + \hat{\mu}_y E_y (t)\right], \]
where the electric field of the $u$-component ($u=x,y$) is defined as the time derivative of vector potential $\vec{A}_u$
\[ \vec{E}_u(t) = - \frac{\partial}{\partial t} \vec{A}_u (t) \]
with
\[A_u (t) = \frac{E_0}{\omega_L} S(t) \sin(\omega_L t - \phi_u). \]
Here, $S(t) = \Theta(t - \tau ) \sin^2(\frac{\pi t}{\tau})$, where $\Theta(t - \tau )$ is the heavy-side function and $\sin^2(\frac{\pi t}{\tau})$ is the envelope of the pulse. It implies,
\[{E}_u(t) = - \frac{E_0}{\omega_L} \Theta (t-\tau) \left(\frac{\pi}{\tau}\right) \sin(\omega_L t - \phi_u) \sin(\frac{2 \pi t}{\tau}) - E_0 S(t) \cos(\omega_L t - \phi_u).
\]
Here, $\omega_L$ is the carrier frequency, $\tau$ is the pulse duration and the phase difference $\phi_u$ determines the polarization.

For left-circularly polarized (LCP) light $\phi_x = 0, \phi_y = \frac{\pi}{2}$; 
and for right-circularly polarized (RCP) light $\phi_x = \frac{\pi}{2}, \phi_y = 0$. The CP light induce transitions subject to the dipole selection rule, $\Delta j = \pm 1$, with $\hat{\mu}_u = \mu_{01} \left( \ket{0}_u \bra{1}_u + h.c \right)$ where $\ket{0}$ and $\ket{1}$ are the Fock states of the u-polarized cavity photon. L(R)CP promotes transitions between $0 \leftrightarrow +1(-1)$ angular momentum sector states. Starting with the initial state of the system in the absolute ground state, denoted by $\ket{\Psi (0)}$ = {$\ket{\{0_{m,1},0_{m,2},..,0_{p}}$}, we apply weak RCP pulse of duration ($\tau$) 20 fs with a carrier frequency ($\omega_L$) 7.24 eV and amplitude $E_0$ = 0.0018 a.u. This pulse prepares a coherent superposition of eigenstates $\ket{\Psi(t)}$ in the sector ($n_{ex} = 1,j = -1$) in the UP branch of the polaritonic spectra for different molecules as shown in Fig. 4 (shaded in green). By solving the time-dependent Schr\"odinger equation (TDSE) for $\Psi(t)$ with the MCTDH method including $\hat{H}_I$, the time-dependent polarizations triggered by the RCP light are obtained by $\braket{\hat P(t)} = \bra{\Psi(t)}\hat P\ket{\Psi(t)}$, where the $\hat P$ is defined in terms of $a_{\pm} ^{(\dagger)}$ in the main text. Within the the weak-field limit, the resulting polarization, normalized by the excited-state population after the pulse, remains unchanged with respect to the intensity of the RCP pulse.

%\bibliography{JT-cavity.bib}
%apsrev4-2.bst 2019-01-14 (MD) hand-edited version of apsrev4-1.bst
%Control: key (0)
%Control: author (8) initials jnrlst
%Control: editor formatted (1) identically to author
%Control: production of article title (0) allowed
%Control: page (0) single
%Control: year (1) truncated
%Control: production of eprint (0) enabled
%